# Low-surface energy carbon allotrope: the case for bcc-C$_6$

Wen-Jin Yin[1,2], Yuan-Ping Chen[1], Yue-E Xie[*1], Li-Min Liu[*2], and S. B. Zhang[3]

[1]Department of Physics, Xiangtan University, Xiangtan 411105, Hunan, China
[2]Beijing Computational Science Research Center, Beijing 100084, China
[3]Rensselaer Polytechnic Institute, Troy, NY 12180, USA

[*1]Corresponding author: xieyech@xtu.edu.cn
[*2]Corresponding author: limin.liu@csrc.ac.cn

**Abstract**

Graphite may be viewed as a low-surface-energy carbon allotrope with little layer-layer interaction. Other low-surface-energy allotropes but with much stronger layer-layer interaction may also exist. Here, we report a first-principles prediction for one of the known carbon allotropes, bcc-C$_6$ (a body centered carbon allotrope with six atoms per primitive unit) that should have exceptionally low-surface energy and little size dependence down to only a couple layer thickness. This unique property may explain the existence of the relatively-high-energy bcc-C$_6$ due to the easy of nucleation during growth. The electronic properties of the few-layer bcc-C$_6$ can also be intriguing: the (111), (110), and (001) thin layers have a direct band gap, an indirect band gap, and metallic character, respectively. The tamed chemical reactivity of the surface is, however, not totally diminished, as lithium-doped (111) thin layers have noticeably increased binding energies of H$_2$ molecules with a maximum storage capacity of 10.8 wt%.

`

## 1. Introduction

Carbon, the fundamental element on the earth, has attracted great attentions from various fields because of its excellent properties and widespread applications. Recently, a number of bulk allotropes, such as Bct-$C_4$[1], Z-carbon[2], M-carbon[3], W-carbon[4], and bcc-$C_8$[5] have been proposed due to their exceptional properties of hardness, semi-conductivity, or transparency.[6-8] In particular, Goresy et al. reported a new hard transparent phase of carbon found in a rock sample from the Popigai impact crater in Russia.[9] Their x-ray results indicated that this new phase is not amorphous but cubic with *Pm-3m* space group symmetry and a lattice constant of 14.7 Å.[9] Theoretical study suggested a new cage-like bcc-$C_6$ structure with six atoms per primitive unit cell. The calculated lattice parameter, d-spacing, and ground-state enthalpy of the structure are in agreement with experiment.[10] In other words, bcc-$C_6$ is more like to the experimental super-dense carbon than the previously proposed supercubane $C_8$ or the $BC_8$ carbon allotropes.[11-13]

In many aspects, carbon allotropes are not only the material of bulk but also the material of nano-scale.[14-20] The best examples are the graphene as a single atomic layer of graphite and diamondoid as a hydrogenated nanoparticle of diamond.[21-23] Experiments have established that graphene has unique electronic properties, such as half-integer quantum Hall effect, high migration rate, and massless transport characteristics.[24-26] The origin of the exceptional electronic properties can be attributed to the peculiar band structure at the Fermi level, namely, the so-called Dirac cores.[24] The T-graphene, obtained by cleaving the bcc-$C_8$ or Bct-$C_4$ along the (001) plane, also possesses different properties from bulk. Although by in large, few studies on the nanostructures of carbon allotropes have been performed.[27] Recently, other two-dimensional layered structures, such as transition metal dichalcogenide (TMD) attracted much attention.[28-31] Among the reasons, the TMD exhibits an indirect to direct transition at monolayer thickness.[28, 30-32] This raises an important question, can the indirect to direct transition take place among the carbon allotropes, which could potentially broaden the carbon-based nanostructures for optoelectronic application?

In this paper, we carry out a systematic study of two-dimensional bcc-$C_6$ using first-principles methods. The results show that thin layers of bcc-$C_6$ could also exhibit exceptional structures and electronic properties. In particular, the surface energy of bcc-$C_6$ is exceptionally low, only 0.1-0.2 eV per C for the (111), (110), and (100) surfaces. The effect of quantum confinement on the energy is also minimal, only up to a couple tenths of an eV per C. The indirect band gap of bulk bcc-$C_6$ becomes direct one for (111) thin layers. At low dimension, the chemistry of each face can also change dramatically. In particular, $H_2$ adsorption energy on Li-doped *N = 3* bcc-$C_6$ (111) thin layers is increased to an average of 0.25 eV/$_{H2}$ with a total hydrogen storage capacity of 10.8 wt%.

## 2. Computational Method

The first-principles calculations based on the density functional theory (DFT) have been carried out by the Vienna ab initio simulation program package (VASP).[33, 34] Exchange-correlation effects were described by the Perdew-Burke-Ernzerhof (PBE) of generalized gradient approximation (GGA) for most of the calculations, while HSE06 functional were used when necessary.[35] The projector augmented wave (PAW) method has been used to represent the atomic cores. An energy cutoff of 500 eV has been set for the plane-wave basis, which ensures that the total energies converge to 0.001eV per atom. The self-consistent filed (SCF) convergence criterion between two ionic steps is set at $10^{-7}$ eV. In some case, the Monkhorst-Pack **k**-point grids are set as 11×11 in xy plane and one in z direction. In order to avoid the interactions between adjacent images, a vacuum region of 15Å is added. The geometries were fully relaxed until the residual forceless than 0.001eV/Å on each atom. To verify the stability of the configurations, the phonon dispersion were calculated by using the Phonopy package.[36]

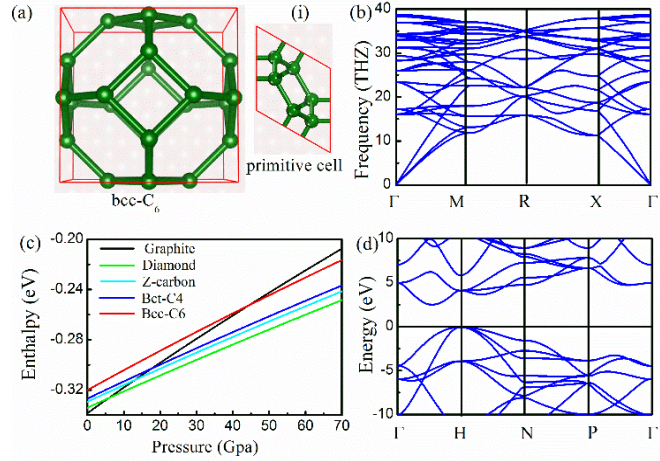

*Fig. 1 (color online). Structure and properties of bulk bcc-$C_6$. (a) A perspective view of the structure consisting of hexagonal and tetrahedral rings forming a $C_{24}$ cage. Inset (i) shows primitive cell with six C atoms. (b) Calculated phonon dispersion for bcc-$C_6$ at zero pressure. (c) Enthalpies versus pressure for graphite, diamond, bcc-$C_6$, and other carbon allotropes. (d) Electronic structure of bcc-$C_6$.*

## 3. Results and Discussion

Let us first consider bulk bcc-$C_6$, which has an *Im-3m* space group with $sp^3$ bonds. The conventional cell of the bcc-$C_6$ consists of tetrahedral ring ($C_4$) and hexagonal ring ($C_6$) forming a $C_{24}$ cage shown in Fig. 1(a). While the primitive cell includes only six atoms occupying the Wickoff position 12*d* (0.5, 0.25, 0.0), together with lattice parameter $a = b = c = 3.8$ Å and volume of $V_0 = 42.14$ Å$^3$ (see insert (i) in Fig. 1(a)). The relaxed C-C bond length is about 1.55 Å, which is close to diamond of 1.54 Å.[37] Bond angles are either 90° in tetrahedral ring or 120° in hexagonal ring. In order to verify the relative stability of bcc-$C_6$, the formation energy is calculated along with the other carbon allotropes such as bcc-$C_8$ and T carbon. It shows that bcc-$C_6$ has a formation energy of -8.70 eV/atom, which is lower than bcc-$C_8$ of -8.49 eV/atom and T carbon of -7.95 eV/atom,[27, 38] indicating that bcc-$C_6$ is relative stable. In addition, thermodynamic meta-stability of bcc-$C_6$ is also verified by phonon vibration frequencies. Figure 1(b) shows that all the phonon spectrum branches are positive and no imaginary phonon mode exists.

The bcc-$C_6$ was found at impact crater[9], which suggests that it may be formed under high

`

temperature and high pressure. Previous LDA result, however, showed that bcc-$C_6$ cannot be obtained by applying a high pressure.[10] Our results agree with the fact that bulk bcc-$C_6$ is generally high in energy. However, its enthalpy at elevated pressure can be lower than that of graphite (see Fig. 1(c)), when the pressure exceeds about 45 GPa. Figure 1(d) shows the electronic structure for bulk bcc-$C_6$. It has an indirect band gap of 2.5 eV, which is much smaller than other carbon allotropes, such as diamond (5.4 eV), Bct-$C_4$ (3.04 eV).[27] The valence band maximum (VBM) is located at H point, while the minimum conduction band (CBM) is between Γ and H point.

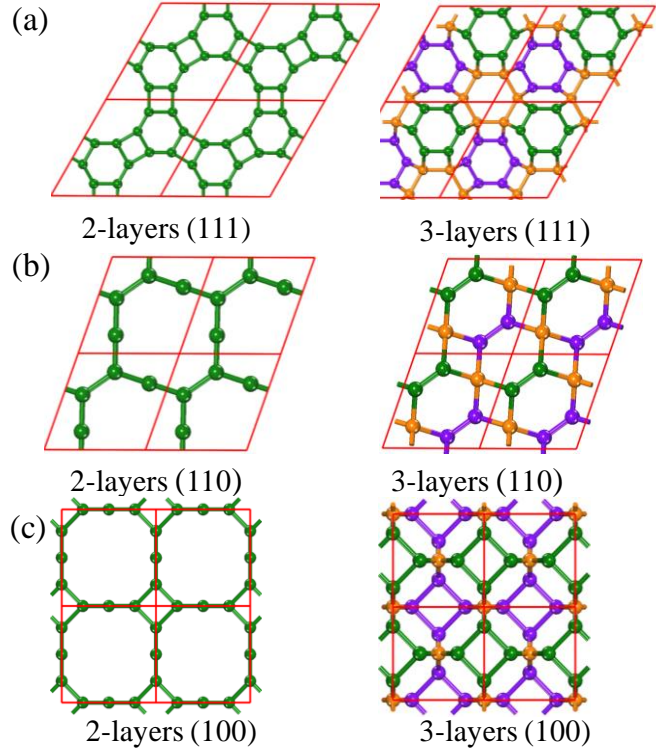

Fig. 2 (color online). Two-dimensional layered bcc-$C_6$ with (a) (111), (b) (110), and (c) (100) surfaces. Atoms in the first, middle, and third layers are shown in green, orange, and purple, respectively.

Two-dimensional materials often exhibit excellent properties over their bulk materials. In order to know whether the low dimensional bcc-$C_6$ also have different properties from bulk, its slabs of three faces [(111), (110), and (100)] with thickness *N* are considered, here *N* is the nominal thickness of the slab. Figure 2 shows the relaxed structures for the slabs with *N = 2* and *3*. One can see that all *N = 2* structures relax into monolayer structures after structural optimization, while all slabs will maintain multiply atomic layers after *N = 3*. The right panel in Fig. 2(a) shows that the slabs of face (111) are composed of multi-layer regular hexagonal rings and the hexagonal rings are connected by rectangle rings. For the slabs of face (100), the right panel in Fig. 2(c) indicates that they are made of tetrahedral rings connected by single carbon atom in the middle layer. While the slabs of face (110) are complicated relatively, where only the structures with *3N* atomic layers are considered to avoid the formation of extensive dangling bonds (see the right panel in Fig. 2(b)).

`

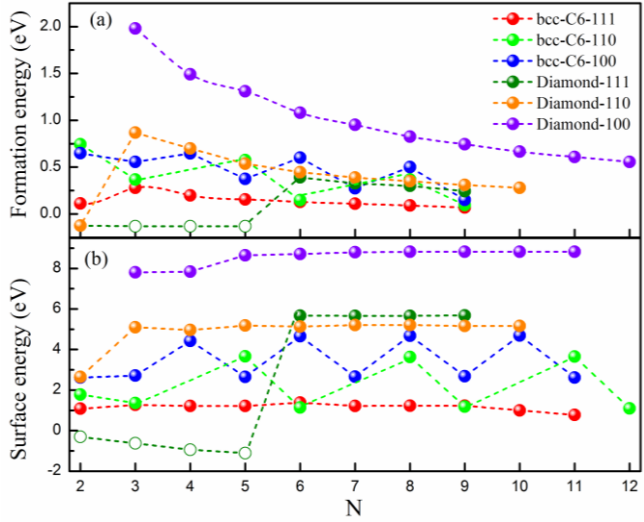

*Fig. 3 (color online). Formation energy and surface energy of the two-dimensional bcc-$C_6$ and diamond in faces (111), (110), and (100) as a function of N. N denotes the number of atomic layers.*

In order to determine the relative stability of these two-dimensional slabs, we calculate the formation energy defined as $E_f = (E_{slab}-E_{bulk})/n$, where $E_{slab}$ is the total energy of the slab calculated in the condition of fully box and atom relaxed in the supercell approximation, and $E_{bulk}$ is the total energy of bulk with the same number of atoms ($n$) in the slab. The calculated results are shown in Fig. 3(a). The formation energy of diamond slabs of faces (111), (110) and (100) are also given for comparison. One can find that the slabs of face (111) have extremely small formation energies even for the very thin slabs, and the formation energies converge to about 0.05 eV with the increase of thickness. While the formation energies of diamond (111)-face slabs converge to about 0.2 eV (the negative formation energies represent the structures are multi-layer graphene after geometry optimization). The formation energies for the bcc-$C_6$ (110)- and (100)-face slabs oscillate with the number of layers because of different surface reconstruction. But most of these values are still lower than those of corresponding diamond structures. Therefore, the bcc-$C_6$ thin slabs exhibit high stability than the diamond slabs. To further explain this point, we calculate the surface energies $E_s$ of bcc-$C_6$ and diamond slabs, as shown in Fig. 3(b). The surface energies are defined as $E_s = (E_{slab}-E_{bulk})/2S$, where $E_{slab}$ is obtained in the condition of only atom relaxed, and $S$ is the surface area. It can be observed that each face of the bcc-$C_6$ has a relative low surface energy than that of diamond. This may explain the existence of the relatively-high-energy bcc-$C_6$ due to the easy of nucleation during growth. Among all faces, the bcc-$C_6$ (111) thin layers have the lowest surface energy, suggesting the easy nucleation on this face. Besides the diamond, the surface energies of other carbon allotropes such as Bct-$C_4$, Z-carbon, M-carbon with N = 12 are further checked and their results are given in Table I. In these carbon allotropes, the bcc-$C_6$ also has the lowest surface energies in various surfaces. For example, the surface energy of bcc-$C_6$ (111) face (0.77 J/m$^2$) is the lowest and the (110) face is the second lowest surface energy (1.09 J/m$^2$). Although the energy of bcc-$C_6$ (100) face (2.62 J/m$^2$) is higher than that of Bct-$C_4$ and Z-carbon, it is still much lower than that of other carbon allotropes.

`

These results further indicate that bcc-$C_6$ has an exceptionally low surface energy. Therefore, it is relatively easy to fabricate them due to the vanishing nucleation barriers despite its relatively large bulk energy.

*Table I.* PBE surface energy (J/m$^2$) of various carbon allotropes. (a) is the LDA results from Ref. [33].

| Carbon allotropes | (111) | (110) | (100) |
|---|---|---|---|
| **Bcc-$C_6$** | **0.77** | **1.09** | **2.62** |
| Bct-$C_4$ | 5.43 | 4.21 | 1.31 |
| M-carbon | 6.24 | 4.98 | 4.20 |
| W-carbon | 4.06 | 5.20 | 2.01 |
| Z-carbon | 5.50 | 4.32 | 1.75 |
| Diamond | 5.66 | 5.16 | 8.83 |
| Diamond[a] | 6.43 | 5.93 | 9.40 |

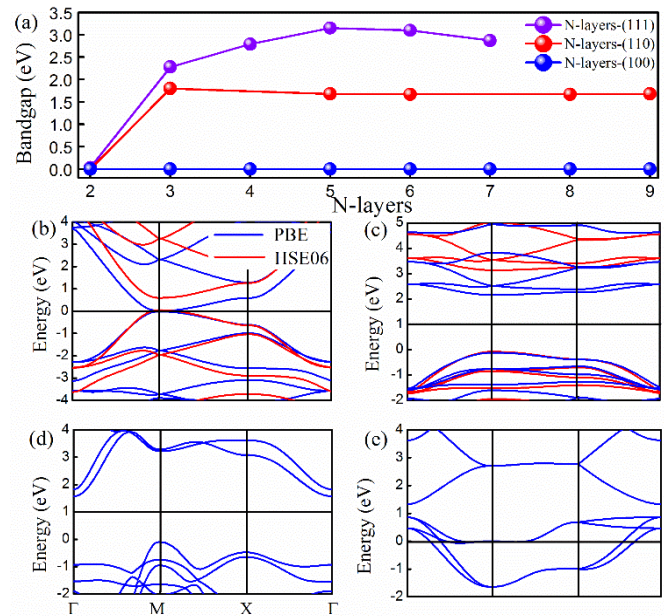

*Fig.4* (color online). Electronic structure for the bcc-$C_6$ thin layers. (a) Band gap as a function of thickness N for each face. Band structures for (b) N = 2 (111) thin layers, and N = 3 (c) (111), (d) (110), and (e) (100) thin layers. In (b)-(c), PBE and HSE06 results are in blue and red lines, respectively.

Figure 4(a) shows the calculated band gap for bcc-$C_6$ thin layers, as a function of thickness. In order to know the detailed electronic properties of these thin layers, the band structures are also given in Figs. 4(b)-(e). It appears that N = 2 is an exception (e.g., 0.02 eV for face (111) in Fig. 4(b)). On the other hand, N = 3 (Figs. 4(c)-(e)) is representative of the band structures for N > 3. The (111) thin layers have a direct band gap at M point and the gap value saturates at about 3.3 eV, much like the model photon-catalysis material $TiO_2$ of 3.2 eV.[39] The (110) thin layers have an indirect band gap of about 1.8 eV, which is independent of N except for N

`

= 2. The VBM is located at M point, whereas the CBM is at Γ point. In stark contrast, the (100) thin layers are always metallic independent of N. Such a metallic behavior may originate from the presence of tetrahedral ring, as having been found in planar T-graphene.[27] PBE calculation usually underestimates the band gap of semiconductors. To improve the accuracy, we have calculated the band gaps of the (111) thin layers using a hybrid functional, HSE06: the N = 2 gap increases from 0.02 to 0.6 eV, while the N = 3 gap increases from 2.2 to 3.2 eV.

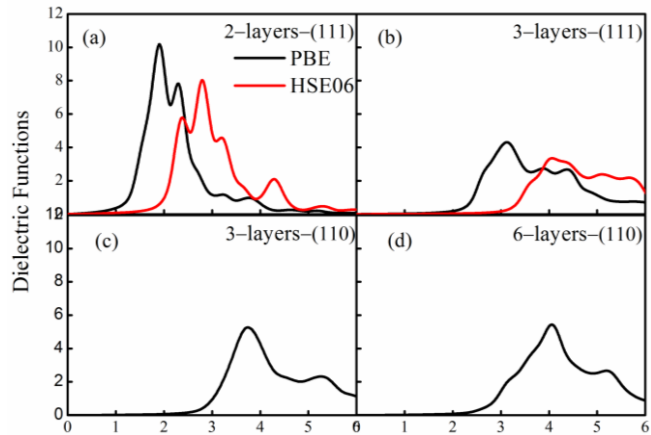

*Fig.5 (color online). Imaginary part of the dielectric function as a function of energy for (a) N = 2 (111), N = 3 (b) (111) and (c) (110), and (d) N = 6 (110) thin layers. PBE and HSE06 are in black and red, respectively.*

Figure 5 shows the optical properties for direct-gap (111) thin layers, along with those for indirect gap (110) thin layers for comparison. Within PBE, the absorption threshold for N = 2 (111) is at about 1 eV, which is noticeably larger than the gap of 0.02 eV. For N = 3, however, significant absorption starts right at the band gap of about 2 eV. Hence, the N = 3 (111) thin layers is indeed a direct gap semiconductor. In contrast, the (110) thin layers all have absorption edges larger than the band gap, as expected for indirect materials. To correct for the band gap errors, HSE06 functional is also used for the (111) thin layers. The absorption threshold for N = 2 is increased from 1 to 2 eV, while for N = 3, it is increased to 3 eV, in line with the gap change from PBE to HSE06.

It has been suggested that Li-doped graphene or porous graphene could have a good $H_2$ storage capacity.[3] Given the buckled carbon structure for N = 3 thin layers, it makes sense to study if Li-doping could also significantly enhance $H_2$ adsorption. Our result shows that Li takes the interstitial site at the center of the hexagonal carbon ring. Figure 6 shows the calculated $H_2$ absorption energy as a function of the number of $H_2$ molecules ($N_{H2}$). It can find that the first $H_2$ prefers to adsorb on top of a carbon atom, as shown in the first inset to the left. As $N_{H2}$ increases, more carbon sites are taken, while maximizing the separation between the adsorbed $H_2$ molecules. A maximum of four $H_2$ molecules can be adsorbed in this fashion. When $N_{H2}$ exceeds, additional $H_2$ molecules assume the hollow sides.

The PBE adsorption energy is 0.31 eV/$H_2$ for the first $H_2$, which is significantly larger than what was reported (0.18 eV/$H_2$) for Li-doped graphene.[40] As $N_{H2}$ increases, the adsorption energy first increases to 0.33 eV/$H_2$, but then decreases. When $N_{H2}$ = 6, the adsorption energy

`

is 0.15 eV/$_{H2}$, which is close to the suggested optimal adsorption energy of 0.155 eV/$_{H2}$ for hydrogen storage.[41] Based on Fig. 6, we estimate the maximum capacity to be six $H_2$ in one unit cell or 10.8 wt%.

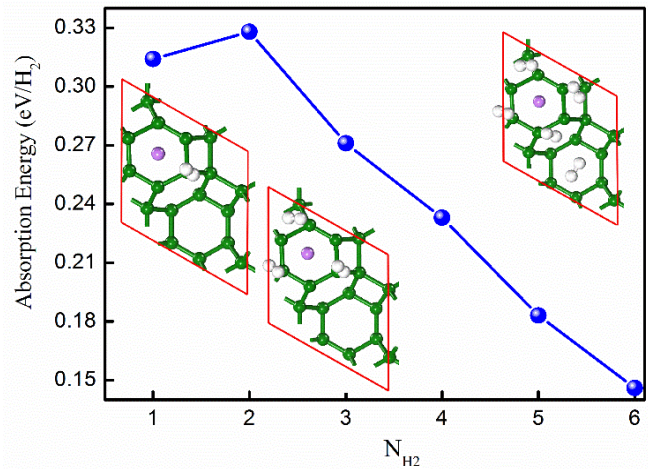

*Fig. 6 (color online). Adsorption energy of $H_2$ on a Li-doped N = 3(111) thin layers, as a function of the number of $H_2$ molecules. Insets show (from left to right) the optimized geometries for 1, 3, and 5 $H_2$. Green, purple, and white balls are C, Li, and H atoms, respectively.*

## 4. Conclusion

First-principles study reveals that bulk bcc-$C_6$ has an indirect band gap of 2.5 eV. It is metastable, as evidenced by the lack of any negative phonon energies. It belongs to a class of interesting materials with exceptionally-low surface energy and small size dependence of its formation energy. Thus, at the nucleation stage when the crystal size is exceedingly small, bcc-$C_6$ could be more stable than other diamond forms to first nucleate. This kinetic stability may explain the experimental observation of the carbon allotrope with very similar properties to bcc-$C_6$. The unique properties of bcc-$C_6$ at nanoscale also make it an excellent material for building few-layer structures. It is interesting to note that the band gap can either switch from indirect to direct, remain indirect, or become metallic properties, when the layered bcc-$C_6$ are (111), (110) and (001) orientated, respectively. Increased chemical affinity of the (111) surface is also predicted. In particular, $H_2$ adsorption energy can be significantly increased to an average value of 0.25 eV/$_{H2}$ on Li-doped structures with a maximum capacity of 10.8 wt%.

## Acknowledgments

This work was supported by the National Natural Science Foundation of China (Nos. 51222212, 51176161, 51376005, 11474243, 11244001 and 51032002), the MOST of China (973 Project, Grant NO. 2011CB922200), the Ministry of Science & Technology of China (Project 2012AA050704). The computations supports from Informalization Construction Project of Chinese Academy of Sciences during the 11th Five-Year Plan Period (No.INFO-115-B01) are also highly acknowledged. SBZ was supported by the Department

`

of Energy under Grant No. DE-SC0002623.

`

`

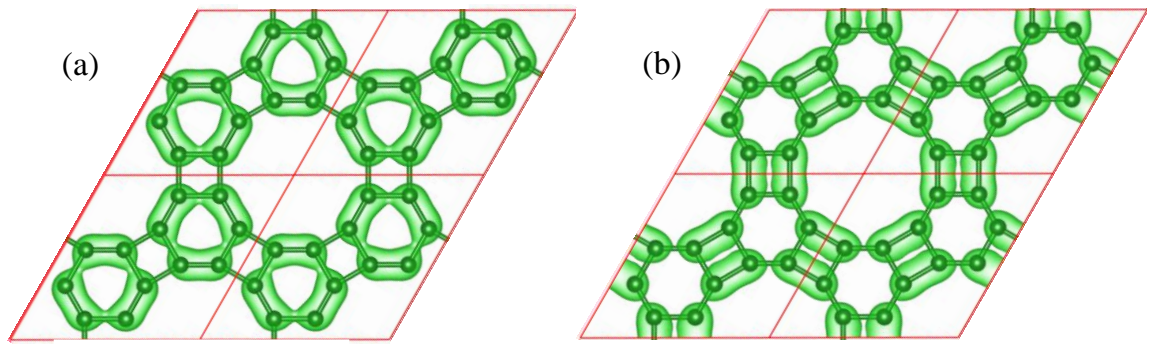

*Fig. S1* Charge density distributions of the bcc-$C_6$ $N = 2$ (111) thin layers, (a) VBM and (b) CBM.

`